\documentclass[final,twocolumn,10pt]{IEEEtran}

\usepackage{spconf}
\usepackage{amsmath,epsfig,amssymb,algorithm,amsthm,cite,url}

\usepackage{algpseudocode,tabularx}
\usepackage{flushend}
\usepackage[shrink=20]{microtype}

\usepackage{tikz,pgfplots}
\usetikzlibrary{arrows,positioning,shapes,backgrounds,fit} 
\usetikzlibrary{arrows.meta} 
\usetikzlibrary{decorations.pathreplacing}
\usetikzlibrary{positioning,calc}
\usetikzlibrary{matrix}

\usepackage{subcaption} 
\usepgfplotslibrary{groupplots}
\pgfplotsset{compat=newest}

\usepackage{rotating}

\usepackage[font=scriptsize]{caption}

\newtheorem*{remark*}{Remark}

\usepackage{tikz,pgfplots}
\usepackage{bm}
\usetikzlibrary{spy}
\pgfplotsset{compat=1.7}
\usetikzlibrary{shapes.geometric, intersections,plotmarks}
\DeclareMathOperator*{\argmin}{\arg\!\min}
\DeclareMathOperator*{\argmax}{\arg\!\max}

\title{Latent heterogeneous multilayer community detection}
\name{Hafiz Tiomoko Ali$^1$, Sijia Liu$^5$, Yasin Yilmaz$^2$, Romain Couillet $^{4}$, Indika Rajapakse$^3 $, Alfred Hero$ ^3 $ \thanks{This work was partially supported by DARPA under the Deep Purple and FunCC program and by ARO under grant W911NF-15-1-0479. Couillet's work is supported by the GSTATS UGA IDEX DataScience chair and the ANR RMT4GRAPH Project (ANR-14-CE28-0006).}}
\address{$^1$ Huawei Noah's Ark Lab, $ ^2$ University of South Florida , $^3$ University of Michigan \\$^4$ GIPSA-lab, University Grenoble-Alpes, $^5$ MIT-IBM Watson AI Lab, IBM Research.}

\begin{document}
\ninept
\maketitle
\begin{abstract}
We propose a method for simultaneously detecting shared and unshared communities in heterogeneous multilayer weighted and undirected networks. The multilayer network is assumed to follow a generative probabilistic model that takes into account the similarities and dissimilarities between the communities. We make use of a variational Bayes approach for jointly inferring the shared and unshared hidden communities from multilayer network observations. We show that our approach outperforms state-of-the-art algorithms in detecting disparate (shared and private) communities on synthetic data as well as on real genome-wide fibroblast proliferation dataset.
\end{abstract}
\section{Introduction}
\label{introduction}
Community detection in networks is an ubiquitous area of active research as it enables the exploration of network structural properties in real-world scenarios including, but not limited to, social and biological sciences ~\cite{fortunato2010community}. There is a vast literature on single-layer network community detection using modularity optimization \cite{newman2006modularity}, spectral clustering \cite{newman2013spectral} and statistical inference \cite{hastings2006community}. However, single-layer networks are not well suited to real-world networks, such as the Internet of Things (IoT), transportation, social, and biological networks where multi-relational, multidimensional, multiplex or multilayer networks structures exist~\cite{boccaletti2014structure}. In social networks, for example, large company employees may be related to each other by similarity of their activities (functional relations) within the organization,  and also by sharing the same office location (spatial relations)~\cite{oselio2014multi}. As another example, in genomics, genes might be related either by transcriptional interactions (functional relations), e.g., measured by RNA-seq profile similarity, or by chromatin interactions (spatial relations), e.g., measured by chromatin conformation capture (Hi-C) of promter-enhancer ligations~\cite{dixon2015chromatin}. In each of these examples, there may be a community structure that is common to each layer and a structure that is distinct between layers. Such multilayer networks create a need for new community detection methods~\cite{cai2005community}. Recovering the communities from each network independently is suboptimal as this strategy does not exploit the shared information across the network layers. Thus, current research efforts aim at developing joint inference methods by multilayer aggregation\cite{de2015identifying}. The simplest aggregation approach is to collapse the multilayer network to a single-layer network on which classical community detection methods can be applied~\cite{tang2009uncoverning,tang2012community,amelio2014community}. Alternatively, some researchers have suggested performing community detection separately in each layer followed by consensus aggregation of the communities across layers ~\cite{oselio2014multi,amelio2014community, paul2016null}. Another approach is to extend single-layer   Stochastic Block Models~\cite{holland1983stochastic} to multilayer networks~\cite{valles2016multilayer,reyes2016stochastic, barbillon2017stochastic}.

In many applications, some communities might be shared between the different layers, while others not (see Figure~\ref{fig:sfig3}). However, few methods in the literature explicitly consider this general scenario. The Multilayer Extraction algorithm proposed in~\cite{wilson2017community} allows for the identification of heterogeneous multilayer communities where the communities might be shared between a subset of layers. The authors in~\cite{wilson2017community} minimize a cost function and take into account similarities and dissimilarities between the layers' communities. While the model used in~\cite{wilson2017community} is realistic when considering multilayer graph connections, the method is only limited to unweighted graphs. The approach proposed in~\cite{boden2012mining} extends~\cite{zeng2006coherent} to weighted multilayer graphs and allows the extraction of coherent dense subgraphs (cliques) shared by subsets of layers. However, those methods are limited to the identification of dense communities and might fail when the communities are connected but with only a few edges. 

In the present article, we propose a new model-based method to simultaneously detect shared and unshared communities between heterogeneous weighted networks. We define joint weighted stochastic block models (WSBM) that share ``a part" of their community structures. We develop a mean field variational Bayes approach to infer the latent shared and private communities from the proposed multilayer WSBM. This extends the works in~\cite{aicher2014learning, zhang2017theoretical} devised for WSBM in single-layer graphs. Our main contributions are
\begin{itemize}
\item We derive a variational Bayes algorithm for automatically inferring shared and unshared communities from multilayer weighted graphs. 
\item We establish that the proposed algorithm is more accurate and robust than previous approaches to community detection in multi-layer networks in extracting both shared and unshared communities from weighted graph benchmarks. 
\item We illustrate a real-world use of our method in multinomic molecular biology where it enables the discovery of heterogeneous multilayer communities of gene-gene interactions in human fibroblast proliferation.
\end{itemize}

{\bf Notations}: Vectors are denoted with boldface lowercase letters and matrices by boldface uppercase letters. $ {\bf I}_n \in \mathbb{R}^{n \times n}$ is the identity matrix and $ {\bf 1}_n $ is the  column vector full of ones.

\section{Joint Weighted Stochastic Block Models}
\label{sec:benchmark}
To start with, let us recall the definition of a single-layer Stochastic Block Model generated by a weighted distribution $\mathcal{D}$ with sufficient statistic $ T $ and natural parameter $ \eta $. Given \emph{latent} community labels $g_i \in \{1,\ldots,K\}$ (with $ K$ denoting the number of communities) of each vertex $ i $ and a community-wise connectivity matrix $ {\bm \theta} \in \mathbb{R}^{K \times K},$ an edge is placed between two vertices $ i $ and $ j $ with an adjacency weight $ A_{ij} $ such that
\begin{align*}
\mathbb{P}(A_{ij}|g_i,g_j,\theta_{g_i,g_j}) &\propto \exp\left\{T(A_{ij})\eta(\theta_{g_ig_j})\right\}.
\end{align*}
Following a Bayesian approach, prior distributions are attributed to the labels $ g_i $ and the community-wise connectivity matrix $ {\bm \theta}.$

We denote a multilayer graph, $ \mathcal{G},$ defining $ L $ as the number of layers and $ n $ as the number of vertices. The graph in the $ l$-th layer is an undirected (possibly weighted) graph $ \mathcal{G}^{(l)}=(\mathcal{V},\mathcal{E}^{(l)}) $ with $ \mathcal{V} $ denoting the set of common vertices and $ \mathcal{E}^{(l)} $ denoting the set of edges in graph $ \mathcal{G}^{(l)}.$ We denote by $ {\bf A}^{(l)} $ the adjacency matrices containing the edge weights between each pair of vertices in graph $ \mathcal{G}^{(l)}$. We propose the following generative heterogeneous community structure of the multilayer graph $ \mathcal{G} $. 
\begin{enumerate}
\item We assume that each layer is subdivided into $ K^{(l)} $ non-overlapping communities among which the first $ K\leq \min_l K^{(l)} $ are shared between the layers as described below.
\item \label{enum:Point2} We first generate the label $ g_i^{(1)} $ of each vertex $ i $ in the first layer as $ g_i^{(1)} \sim $ Categorical $({\bm \mu}_0^{(1)}),$ where $ {\bm \mu}_0^{(1)} \in \mathbb{R}^{K^{(1)}}$ contains prior probabilities that the vertices belong to one of the $ K^{(1)} $ communities.
\item \label{enum:Point3} For each vertex $ i,$ if $ g_i^{(1)} \in \{1,\ldots,K\} $ then set $  g_i^{(l)}=g_i^{(1)}$ for each layer $ l.$ Otherwise, generate for each layer $ l,$ $ g_i^{(l)} \sim $ Categorical $({\bm \mu}_0^{(l)}).$
\item Given \emph{latent} community labels $g_i^{(l)}$ (generated in steps~\ref{enum:Point2} and~\ref{enum:Point3}) of each vertex $ i $ and community-wise connectivity matrices $ {\bm \theta}^{(l)} \in \mathbb{R}^{K^{(l)} \times K^{(l)}}$ (the generation of which will be defined later), an edge is placed between two vertices $ i $ and $ j $ and is assigned an adjacency weight $ A^{(l)} _{ij} $ drawn according to 
\begin{align}
\label{eq:likelihood}
\mathbb{P}(A^{(l)}_{ij}|g_i^{(l)},g_j^{(l)},\theta^{(l)}_{g^{(l)}_ig^{(l)}_j}) &\propto \exp\left\{T^{(l)}(A^{(l)}_{ij})\eta^{(l)}(\theta^{(l)}_{g^{(l)}_ig^{(l)}_j})\right\},
\end{align}
where $ T^{(l)} $ is the sufficient statistic and $ \eta^{(l)}$ is the natural parameter of the weights distribution.
\item The community-wise connectivity matrices $ {\bm \theta}^{(l)} $ are generated according to conjugate priors associated with the distribution characterized by $(T^{(l)}, \eta^{(l)})$ , i.e.,
\begin{align*}
p^{\star}(\theta^{(l)}_{ab})&=\frac{1}{Z^{(l)}(\tau_0^{(l)})}\exp(\tau_0^{(l)}\eta^{(l)}(\theta^{(l)}_{ab}))
\end{align*}
with $ \tau_0^{(l)}$ denoting the associated hyperparameters and $ Z^{(l)}(\tau_0^{(l)})$ the normalization constants.
\end{enumerate}

For illustration, we specialize the presentation to two communities, for which each of the matrices $ {\bm \theta}^{(l)} $ are decomposed into four blocks corresponding, respectively, to the shared-shared, shared-private, private-shared, private-private interconnections (see Figure~\ref{fig:sfig4}). As in~\cite{aicher2014learning}, we consider each sub-matrix ${\bm \theta}^{(l)}_1, {\bm \theta}^{(l)}_2, {\bm \theta}^{(l)}_3, {\bm \theta}^{(l)}_4$ as one-dimensional vectors where the elements are stacked. Let us denote by $ r^{(l)}_1, r^{(l)}_2, r^{(l)}_3, r^{(l)}_4 $ the indexing variables into each of the obtained vectors , i.e., $ r^{(l)}_1=1,\ldots,K^2; r^{(l)}_2=1,\ldots,K(K^{(l)}-K); r^{(l)}_3=1,\ldots,K(K^{(l)}-K); r^{(l)}_4=1,\ldots,(K^{(l)}-K)^2.$
\begin{figure}
\centering
  \includegraphics[width=.65\linewidth, height=.4\linewidth]{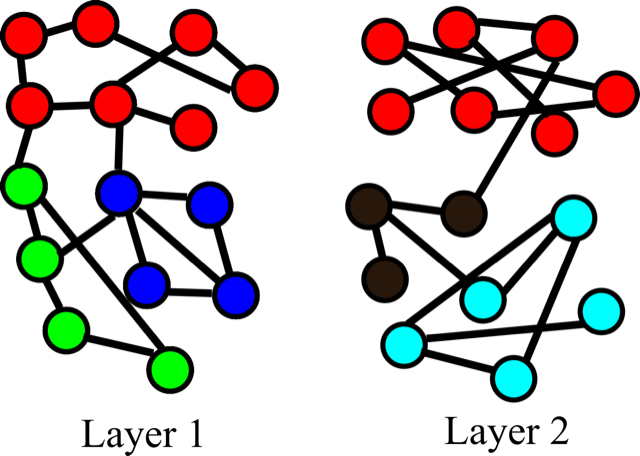}
 \caption{Heterogeneous multilayer network. Shared communities (in red) and unshared communities in different colors for each layer.}
  \label{fig:sfig3}
\end{figure}
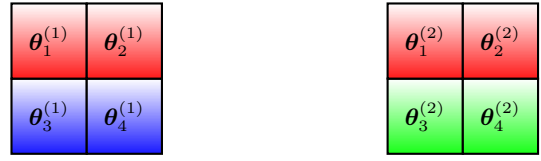
\begin{figure}
\begin{tikzpicture}[scale=1]
\filldraw[thick, top color=white,bottom color=blue!90!] (0,-3) rectangle node{${\bm \theta}^{(1)}_3$} +(1,1);
\filldraw[thick, top color=white,bottom color=blue!90!] (1,-3) rectangle node{${\bm \theta}^{(1)}_4$} +(1,1);
\filldraw[thick, top color=white,bottom color=red!90!] (0,-2) rectangle node{${\bm \theta}^{(1)}_1$} +(1,1);
\filldraw[thick, top color=white,bottom color=red!90!] (1,-2) rectangle node{${\bm \theta}^{(1)}_2$} +(1,1);

\filldraw[thick, top color=white,bottom color=green!90!] (5,-3) rectangle node{${\bm \theta}^{(2)}_3$} +(1,1);
\filldraw[thick, top color=white,bottom color=green!90!] (6,-3) rectangle node{${\bm \theta}^{(2)}_4$} +(1,1);
\filldraw[thick, top color=white,bottom color=red!90!] (5,-2) rectangle node{${\bm \theta}^{(2)}_1$} +(1,1);
\filldraw[thick, top color=white,bottom color=red!90!] (6,-2) rectangle node{${\bm \theta}^{(2)}_2$} +(1,1);


\end{tikzpicture}
\caption{Community affinity probability matrices $ {\bm \theta}^{(1)} $ and $ {\bm \theta}^{(2)} $ decomposed into blocks. Red blocks (common to ${\bf A}^{(1)}$ and ${\bf A}^{(2)}$) used to update shared communities while blue blocks used to update private communities of ${\bf A}^{(1)}$ and green blocks for the update of ${\bf A}^{(2)}$'s private communities.}
\label{fig:sfig4}
\end{figure}%
The overall prior distribution can thus be written as 
\begin{align}
&p^{\star}({\bf g}^{(l)}, {\bm \theta}^{(l)}, l=1,\ldots,L)= \prod_{l=1}^L \prod_i (\mu_0^{(l)})_{i,g^{(l)}_i} \nonumber \\ &\times \prod_{r^{(l)}} \frac{1}{Z^{(l)}(\tau^{(l)}_0)}\exp(\tau^{(l)}_0\eta^{(l)}(\theta^{(l)}_{r^{(l)}}))
\end{align}
with $ r^{(l)}\equiv \{r^{(l)}_1,r^{(l)}_2,r^{(l)}_2,r^{(l)}_4\}.$

Given $\mathcal{G}^{(l)}, l=1,\ldots,L$ (equivalently their adjacency matrices $ {\bf A}^{(l)} $), the goal is to infer the community labels $ g_i^{(l)} $ for each node $ i$ in each layer $ l$ , i.e., to find the most probable clustering $ {\bf g}^{(l)} $ of the vertices in each layer in the set of all different possible partitionings
\begin{align}
\label{eq:prob1}
[({\bf g}^{(1)})^{\star},\ldots,({\bf g}^{(l)})^{\star}]&=\argmax_{{\bf g}^{(l)}, l=1,\ldots,L} \mathbb{P}({\bf g}^{(l)}|{\bf A}^{(l)}, {\bm \theta}^{(l)}, l=1,\ldots,L)
\end{align}
with the correlations constraints on $ {\bf g}^{(l)} $ defined in Point~\ref{enum:Point3}) of Section~\ref{sec:benchmark}. The optimization problem~\eqref{eq:prob1} is NP-hard due to two main difficulties: the maximization is over all possible configurations of $ {\bf g}^{(l)},$ and the calculation of the posterior distribution $ \mathbb{P}({\bf g}^{(l)}|{\bf A}^{(l)},{\bm \theta}^{(l)}),$ which is intractable due to its high dimensional integral form. Our approach to the optimization~\eqref{eq:prob1} is the  mean field variational Bayes approximation~\cite{jordan1999introduction, blei2006variational} that uses a factorisable distribution as an approximation to the joint posterior $p({\bf g}^{(l)},{\bm \theta}^{(l)})\equiv\mathbb{P}({\bf g}^{(l)},{\bm \theta}^{(l)}|{\bf A}^{(l)})$.
\section{Variational inference}
\subsection{Mean field variational Bayes inference}
Denote by $ q({\bf g}^{(l)},{\bm \theta^{(l)}}) $ an approximating (factorisable) distribution that depends on tunable shaping parameters ${\bm \mu}^{(l)}$ and ${\bm \tau}^{(l)}$.
The variational Bayes algorithm fits the distribution $q$ to the joint distribution by minimizing the KL-divergence , i.e., $ q= \argmin_r D_{KL}(r||p).$

Here the distribution $ q $ is taken to have the same parametric form as the prior $ p^{\star} $
\begin{align}
q({\bf g}^{(l)}, {\bm \theta}^{(l)})&=\prod_{l=1}^L \prod_i \mu_{i,g^{(l)}_i}^{(l)}\prod_{r^{(l)}} \frac{1}{Z^{(l)}(\tau^{(l)}_{r^{(l)}})}\exp(\tau^{(l)}_{r^{(l)}}\eta^{(l)}(\theta^{(l)}_{r^{(l)}})),
\end{align}
where $ \tau_{r^{(l)}}^{(l)}$ and ${\bm \mu}^{(l)} \in \mathbb{R}^{n\times K^{(l)}}$ are variational parameters corresponding to the random variables $\theta^{(l)}_{r^{(l)}},$ $  {\bf g}^{(l)}$ respectively. We can rewrite the original problem~\eqref{eq:prob1} as follows
\begin{align*}
[({\bf g^{(1)}})^{\star},\cdots,({\bf g^{(L)}})^{\star}]&=\argmax_{{\bf g^{(l)}}} \int \mathbb{P}({\bf g^{(l)}},{\bm \theta^{(l)}}|{\bf A}^{(l)})\mathrm{d}{\bm \theta^{(l)}} \\
&\approx \argmax_{{\bf g^{(l)}}} \int q({\bf g^{(l)}},{\bm \theta^{(l)}})\mathrm{d}{\bm \theta^{(l)}} \\
&= \argmax_{{\bf g^{(l)}}} \int \prod_l \prod_i q({g_i^{(l)}})q({\bm \theta^{(l)}})\mathrm{d}{\bm \theta^{(l)}} \\
&= \argmax_{{\bf g^{(l)}}}\prod_l \prod_i q({g_i^{(l)}}).
\end{align*}
Since $ q^{(l)} $ is a categorical distribution with parameter $ {\bm \mu}^{(l)},$ the original problem~\eqref{eq:prob1} is equivalent to 
\begin{align}
\label{eq:MAP}
(g^{(l)}_i)^{\star}&=\argmax_{k} {\bm \mu}_{ik}^{(l)}
\end{align}
for each node $ i $ and layer $ l$ and thus the multilayer community detection boils down to a Maximum A Posteriori (MAP) estimator on each nodal variational parameter ${\bm \mu_i^{(l)}}$ for each layer $ l.$ 

\subsection{Learning}
As per~\cite{aicher2014learning}, the constant model likelihood can be written as $\log \mathbb{P}({\bf A}^{(l)})=\mathcal{G}(q)+ D_{KL}(q||p)$
with 
\begin{align}
\label{eq: Gq}
\mathcal{G}(q)&=\mathbb{E}_q \log \mathbb{P}({\bf A}^{(l)}|{\bf g}^{(l)}, {\bm \theta}^{(l)}, l=1,2)+\mathbb{E}_q \frac{p^{\star}}{q},
\end{align}
where $ p^{\star} $ is the prior distribution assigned to the parameters $ {\bf g}^{(l)}, {\bm \theta}^{(l)}.$ Since the likelihood is constant, minimizing $ D_{KL}(q||p) $ (and thus making the approximation $ q $ to be the closest to the sought posterior $ p $) is equivalent to maximizing $ \mathcal{G}(q)$ over the variational parameters.  In the sequel, we devise a procedure to learn the parameters for which $ \mathcal{G}(q)$ is maximized.

We next address how to find the variational parameters $ \tau^{(l)},$ $ {\bm \mu}^{(l)} $ for which $ \mathcal{G}(q) $ is maximized. To this end, let us first compute $ \mathcal{G}(q) $ with the forms of the prior $ p^{\star} $ and the approximation $ q $ defined in the previous section. For illustration, we specialize to $ L=2 $ but the same principle applies to any number of layers.
We have 
\begin{align}
\label{eq: 88}
\mathcal{G}(q)&=\mathbb{E}_q \log \mathbb{P}({\bf A}^{(1)},{\bf A}^{(2)}|{\bf g}^{(1)}, {\bf g}^{(2)}, {\bm \theta}^{(1)}, {\bm \theta}^{(2)})+\mathbb{E}_q \frac{p^{\star}}{q} \nonumber\\
&=\mathbb{E}_q \log \mathbb{P}({\bf A}^{(1)}|{\bf g}^{(1)}, {\bf g}^{(2)}, {\bm \theta}^{(1)}) \nonumber \\ &+\mathbb{E}_q \log \mathbb{P}({\bf A}^{(2)}|{\bf g}^{(1)}, {\bf g}^{(2)}, {\bm \theta}^{(2)})+\mathbb{E}_q \frac{p^{\star}}{q},
\end{align}
where in the last line, we use the chain rule along with the conditional independence between $ {\bf A}^{(1)} $ and $ {\bf A}^{(2)} $ given $ {\bf g}^{(1)}, {\bf g}^{(2)}, {\bm \theta}^{(1)}, {\bm \theta}^{(2)}.$ The structure of the heterogeneous Joint Stochastic Block Model (Section~\ref{sec:benchmark}) couples the random variables ${\bf g}^{(1)}$ and ${\bf g}^{(2)}$ in a simple manner that can be decomposed into the following cases, which we call the \emph{dependency cases}:
\begin{itemize}
\item For a vertex pair $ (i,j) $ belonging to a block with $ {\bm \theta}_1^{(l)},$ $ g^{(1)}_i=g^{(2)}_i $ and $ g^{(1)}_j=g^{(2)}_j.$
\item For a vertex pair $ (i,j) $ belonging to a block with $ {\bm \theta}_2^{(l)},$ $ g^{(1)}_i=g^{(2)}_i $ and $ g^{(1)}_j\neq g^{(2)}_j.$
\item For a vertex pair $ (i,j) $ belonging to a block with $ {\bm \theta}_3^{(l)},$ $ g^{(1)}_i\neq g^{(2)}_i $ and $ g^{(1)}_j=g^{(2)}_j.$
\item For a vertex pair $ (i,j) $ belonging to a block with $ {\bm \theta}_4^{(l)},$ $ g^{(1)}_i\neq g^{(2)}_i $ and $ g^{(1)}_j\neq g^{(2)}_j.$
\end{itemize}

Using these \emph{dependency cases} with~\eqref{eq:likelihood}, we obtain an expression for $ \mathcal{G}(q) $ as per~\eqref{eq: 88}. After differentiation with respect to the sought variational parameters, we obtain updates for $ {\bf \tau}^{(l)}, $ $ {\bm \mu}^{(l)},$ which are stationary points of $ \mathcal{G}(q) $ and correspond to local maxima. The precision of the local maxima depend on the initial values for $ {\bm \mu}^{(l)} .$ A single run of a single-layer clustering algorithm shall lead to the optimal solution basin of attraction. Due to the \emph{dependency cases}, the community memberships variational parameters $\mu_{ik}^{(l)}$ depends on $g^{(l)}_i$ either belonging to the set of shared communities $ \{1,\ldots,K\}, $ or to the set of unshared communities ($\{K+1,\ldots,K^{(l)}\}$).  




Algorithm~\ref{alg:example2} provides the necessary equations for the updates of the variational parameters $ {\bf \tau}^{(l)}$ and ${\bm \mu^{(l)}}.$ Due to~ Equation~\eqref{eq:MAP}, a max decision rule can then be used on $ {\bm \mu^{(l)}} $ to assign labels to each node, namely $ \argmax_k \mu^{(l)}_{ik} $ gives the label assigned to node $ i $ in graph $ \mathcal{G}^{{(l)}}.$ The label of node $ i $ is shared between different graphs $ \{\mathcal{G}^{{(l)}}\}$ when $ \argmax_k \mu^{(l)}_{ik} \in \{1,\ldots,K\}$ and the label is unshared otherwise. 

Algorithm~\ref{alg:example2} is an extension of the variational Bayes algorithm for inferring hidden communities from single-layer graphs~\cite{aicher2014learning,zhang2017theoretical} to the inference of hidden \emph{shared} and \emph{unshared} communities from multilayer graphs. In Algorithm~\ref{alg:example2}, the updates for the parameters $ {\tau}^{(l)} $ are done independently for each graph as in~\cite{aicher2014learning}. As for the community membership variational parameters $ {\bm \mu}^{(l)} \in \mathbb{R}^{n \times K^{(l)}},$ the updates of the first $ K $ columns of $ {\bm \mu}^{(l)} $ are identical and are computed by adding the contributions of each graph. The last $ K^{(l)}-K $ columns of $ {\bm \mu}^{(l)} $ are updated independently using only the information about each graph. This is quite intuitive since the first $ K $ columns of $ {\bm \mu}^{(l)} $ correspond to the shared community evidences and thus they should be updated using the contributions of the graphs altogether, while the last columns correspond to unshared communities and thus the updates should be done independently for each graph.
 
  {\small
\begin{algorithm}[tb]
   \caption{Mean field inference of heterogeneous communities in multilayer graphs.}
   \label{alg:example2}
\begin{algorithmic}
   \State {\bf Inputs:} For $ l=1,\ldots, L,$ layers adjacencies ${\bf A}^{(l)},$ layer distributions $ \mathcal{D}^{(l)}=[T^{(l)}, \eta^{(l)}, Z^{(l)}],$ number of shared communities $ K $, total number $ K^{(l)}$ of communities.
   \State {\bf Output:} $ {\bm \mu}^{(1)}, \ldots, {\bm \mu}^{(L)}.$
      \State For $ l=1,\ldots, L,$ initialize $ {\bm \mu}^{(l)}$ and choose hyperparameters $ \tau_0^{(l)}.$
\Repeat
\For{$l=1$ {\bfseries to} $L$}
\For{$r^{(l)}=1$ {\bfseries to} $(K^{(l)})^2$}
   \State $$ \tau^{(l)}_{r^{(l)}}=\tau_0^{(l)}+\sum_{ij}\sum_{(g^{(l)}_i,g^{(l)}_j)=r^{{(l)}}}T^{(l)}(A^{(l)}_{ij})\mu_{i,g^{(l)}_i}^{(l)}\mu_{j,g^{(l)}_j}^{(l)}.$$
   \EndFor
\EndFor
	\Repeat
	\For{$ i=1 $ {\bfseries to} $n$}
		\For{$ k=1 $ {\bfseries to} $K$}
			\State \begin{align*}
			\mu_{ik}^{(1)}&= \exp \Big\{\frac1L \sum_{l=1}^L \Big[\sum_{r_1^{(l)}, j\neq i \atop (k,g^{(l)}_j)=r_1^{(l)}}T^{(l)}(A^{(l)}_{ij})\mu_{j,g^{(l)}_j}^{(l)}\bar{\eta}^{(l)}_{r_1^{(l)}} \\ &+\sum_{r_2^{(l)}, j\neq i \atop (k,g^{(l)}_j)=r_2^{(l)}}T^{(l)}(A^{(l)}_{ij})\mu_{j,g^{(l)}_j}^{(l)}\bar{\eta}^{(l)}_{r_2^{(l)}}\Big]\Big\} \\
            \bar{\eta}^{{(l)}}_{r^{{(l)}}}&=\frac{\partial \log Z^{{(l)}}}{\partial \tau^{{(l)}}}\Bigr|_{\substack{\tau^{{(l)}}=r^{(l)}}}
			\end{align*}
		\EndFor
\For{$l=1$ {\bfseries to} $L$}
	\State $ \mu_{i,1:K}^{(l)}=\mu_{i,1:K}^{(1)}.$
		\For{$ k=K+1 $ {\bfseries to} $K^{(l)}$}
			\State \begin{align*}
			\mu_{ik}^{(l)}&= \exp \Big\{\sum_{r_3^{(l)}, j\neq i \atop (k,g^{(l)}_j)=r_3^{(l)}}T^{(l)}(A^{(l)}_{ij})\mu_{j,g^{(l)}_j}^{(l)}\bar{\eta}^{(l)}_{r_3^{(l)}} \\ &+\sum_{r_4^{(l)}, j\neq i \atop (k,g^{(l)}_j)=r_4^{(l)}}T^{(l)}(A^{(l)}_{ij})\mu_{j,g^{(l)}_j}^{(l)}\bar{\eta}^{(l)}_{r_4^{(l)}}\Big\}
			\end{align*}
		\EndFor
		\State For $k=1,\ldots, K^{(l)},$ $ \mu_{i,k}^{(l)}=\mu_{i,k}^{(l)}/ \sum_{k=1}^{K^{(l)}} \mu_{i,k}^{(l)}. $
	\EndFor 
\EndFor
	\Until{convergence}
\Until{convergence}
\end{algorithmic}
\end{algorithm}
}

\section{Experiments}
\subsection{Synthetic graphs}
\begin{figure}
\begin{tikzpicture}
   \begin{groupplot}[group style={
                      group name=myplot,
                      group size= 2 by 1},height=4cm,width=4.75cm]
        \nextgroupplot[xmajorgrids,ymajorgrids,
               enlarge x limits=false, xmin=0.3,
      ymin=0.2,
      xmax=0.5,
      ymax=1,
      grid=major,
      scaled ticks=true,
      ylabel={NMI},
      xlabel={$q'$}]
                \addplot[blue, mark=triangle, mark size=2pt, smooth,error bars/.cd,y dir=both,y explicit, error bar style={mark size=.5pt}] plot coordinates{
(0.200000,0.999880)+-(0.000477,0.000477)(0.205000,0.999900)+-(0.000438,0.000438)(0.210000,0.999800)+-(0.000725,0.000725)(0.215000,0.999720)+-(0.000805,0.000805)(0.220000,0.999660)+-(0.000945,0.000945)(0.225000,0.999740)+-(0.000733,0.000733)(0.230000,0.999480)+-(0.001010,0.001010)(0.235000,0.999280)+-(0.001897,0.001897)(0.240000,0.999140)+-(0.001457,0.001457)(0.245000,0.999080)+-(0.001405,0.001405)(0.250000,0.998600)+-(0.002878,0.002878)(0.255000,0.998880)+-(0.003367,0.003367)(0.260000,0.998740)+-(0.001721,0.001721)(0.265000,0.997640)+-(0.004094,0.004094)(0.270000,0.997880)+-(0.003780,0.003780)(0.275000,0.997300)+-(0.003183,0.003183)(0.280000,0.997220)+-(0.003386,0.003386)(0.285000,0.996660)+-(0.002858,0.002858)(0.290000,0.995700)+-(0.004152,0.004152)(0.295000,0.994900)+-(0.005426,0.005426)(0.300000,0.993160)+-(0.006377,0.006377)(0.305000,0.993920)+-(0.004357,0.004357)(0.310000,0.992300)+-(0.005794,0.005794)(0.315000,0.991380)+-(0.005865,0.005865)(0.320000,0.989060)+-(0.007432,0.007432)(0.325000,0.987500)+-(0.008097,0.008097)(0.330000,0.985020)+-(0.012580,0.012580)(0.335000,0.982880)+-(0.011773,0.011773)(0.340000,0.982820)+-(0.008146,0.008146)(0.345000,0.976420)+-(0.017311,0.017311)(0.350000,0.974600)+-(0.012232,0.012232)(0.355000,0.970940)+-(0.013896,0.013896)(0.360000,0.966520)+-(0.018789,0.018789)(0.365000,0.966520)+-(0.017683,0.017683)(0.370000,0.958900)+-(0.018689,0.018689)(0.375000,0.957400)+-(0.017178,0.017178)(0.380000,0.950780)+-(0.019594,0.019594)(0.385000,0.946000)+-(0.021309,0.021309)(0.390000,0.938700)+-(0.020952,0.020952)(0.395000,0.935500)+-(0.023109,0.023109)(0.400000,0.925700)+-(0.019771,0.019771)(0.405000,0.912120)+-(0.029754,0.029754)(0.410000,0.906520)+-(0.027230,0.027230)(0.415000,0.897300)+-(0.025235,0.025235)(0.420000,0.878880)+-(0.032586,0.032586)(0.425000,0.876780)+-(0.024107,0.024107)(0.430000,0.862140)+-(0.023037,0.023037)(0.435000,0.852660)+-(0.028220,0.028220)(0.440000,0.837440)+-(0.027279,0.027279)(0.445000,0.812260)+-(0.032957,0.032957)(0.450000,0.795800)+-(0.033550,0.033550)(0.455000,0.780180)+-(0.039257,0.039257)(0.460000,0.752120)+-(0.041667,0.041667)(0.465000,0.717460)+-(0.046326,0.046326)(0.470000,0.683480)+-(0.070551,0.070551)(0.475000,0.628680)+-(0.094502,0.094502)(0.480000,0.576860)+-(0.109101,0.109101)(0.485000,0.481620)+-(0.101635,0.101635)(0.490000,0.422440)+-(0.079176,0.079176)(0.495000,0.374760)+-(0.064928,0.064928)(0.500000,0.350880)+-(0.061291,0.061291)
};\label{Spectral clustering}
\addplot[red, mark=square, mark size=2pt, smooth,error bars/.cd,y dir=both,y explicit, error bar style={mark size=.5pt}] plot coordinates{
(0.200000,1.000000)+-(0.000000,0.000000)(0.205000,1.000000)+-(0.000000,0.000000)(0.210000,1.000000)+-(0.000000,0.000000)(0.215000,1.000000)+-(0.000000,0.000000)(0.220000,1.000000)+-(0.000000,0.000000)(0.225000,1.000000)+-(0.000000,0.000000)(0.230000,1.000000)+-(0.000000,0.000000)(0.235000,1.000000)+-(0.000000,0.000000)(0.240000,1.000000)+-(0.000000,0.000000)(0.245000,1.000000)+-(0.000000,0.000000)(0.250000,1.000000)+-(0.000000,0.000000)(0.255000,1.000000)+-(0.000000,0.000000)(0.260000,1.000000)+-(0.000000,0.000000)(0.265000,1.000000)+-(0.000000,0.000000)(0.270000,1.000000)+-(0.000000,0.000000)(0.275000,1.000000)+-(0.000000,0.000000)(0.280000,1.000000)+-(0.000000,0.000000)(0.285000,1.000000)+-(0.000000,0.000000)(0.290000,1.000000)+-(0.000000,0.000000)(0.295000,1.000000)+-(0.000000,0.000000)(0.300000,1.000000)+-(0.000000,0.000000)(0.305000,1.000000)+-(0.000000,0.000000)(0.310000,0.999980)+-(0.000200,0.000200)(0.315000,0.999960)+-(0.000281,0.000281)(0.320000,0.999980)+-(0.000200,0.000200)(0.325000,1.000000)+-(0.000000,0.000000)(0.330000,0.999880)+-(0.000686,0.000686)(0.335000,0.999840)+-(0.000677,0.000677)(0.340000,0.999900)+-(0.000438,0.000438)(0.345000,0.999380)+-(0.005011,0.005011)(0.350000,0.999660)+-(0.000807,0.000807)(0.355000,0.999500)+-(0.001150,0.001150)(0.360000,0.999120)+-(0.002280,0.002280)(0.365000,0.998880)+-(0.003859,0.003859)(0.370000,0.998420)+-(0.003319,0.003319)(0.375000,0.998060)+-(0.006706,0.006706)(0.380000,0.996560)+-(0.009416,0.009416)(0.385000,0.995120)+-(0.008137,0.008137)(0.390000,0.993760)+-(0.010773,0.010773)(0.395000,0.991620)+-(0.019296,0.019296)(0.400000,0.989800)+-(0.010287,0.010287)(0.405000,0.980340)+-(0.027621,0.027621)(0.410000,0.978420)+-(0.023906,0.023906)(0.415000,0.972880)+-(0.020917,0.020917)(0.420000,0.959360)+-(0.036665,0.036665)(0.425000,0.957720)+-(0.020763,0.020763)(0.430000,0.944660)+-(0.023101,0.023101)(0.435000,0.934000)+-(0.033702,0.033702)(0.440000,0.918820)+-(0.028703,0.028703)(0.445000,0.888100)+-(0.042046,0.042046)(0.450000,0.866540)+-(0.048380,0.048380)(0.455000,0.852040)+-(0.050932,0.050932)(0.460000,0.814840)+-(0.058184,0.058184)(0.465000,0.771020)+-(0.060217,0.060217)(0.470000,0.726840)+-(0.086630,0.086630)(0.475000,0.660920)+-(0.114507,0.114507)(0.480000,0.599840)+-(0.122293,0.122293)(0.485000,0.486680)+-(0.117478,0.117478)(0.490000,0.421400)+-(0.087691,0.087691)(0.495000,0.367860)+-(0.063488,0.063488)(0.500000,0.343900)+-(0.059998,0.059998)
};\label{Independent mean field}
\addplot[green, mark=diamond, mark size=2pt, smooth,error bars/.cd,y dir=both,y explicit, error bar style={mark size=.5pt}] plot coordinates{
(0.200000,1.000000)+-(0.000000,0.000000)(0.205000,1.000000)+-(0.000000,0.000000)(0.210000,1.000000)+-(0.000000,0.000000)(0.215000,1.000000)+-(0.000000,0.000000)(0.220000,1.000000)+-(0.000000,0.000000)(0.225000,1.000000)+-(0.000000,0.000000)(0.230000,1.000000)+-(0.000000,0.000000)(0.235000,1.000000)+-(0.000000,0.000000)(0.240000,1.000000)+-(0.000000,0.000000)(0.245000,1.000000)+-(0.000000,0.000000)(0.250000,1.000000)+-(0.000000,0.000000)(0.255000,1.000000)+-(0.000000,0.000000)(0.260000,1.000000)+-(0.000000,0.000000)(0.265000,1.000000)+-(0.000000,0.000000)(0.270000,1.000000)+-(0.000000,0.000000)(0.275000,1.000000)+-(0.000000,0.000000)(0.280000,1.000000)+-(0.000000,0.000000)(0.285000,1.000000)+-(0.000000,0.000000)(0.290000,1.000000)+-(0.000000,0.000000)(0.295000,1.000000)+-(0.000000,0.000000)(0.300000,1.000000)+-(0.000000,0.000000)(0.305000,1.000000)+-(0.000000,0.000000)(0.310000,0.999980)+-(0.000200,0.000200)(0.315000,0.999960)+-(0.000281,0.000281)(0.320000,0.999980)+-(0.000200,0.000200)(0.325000,1.000000)+-(0.000000,0.000000)(0.330000,0.999880)+-(0.000686,0.000686)(0.335000,0.999840)+-(0.000677,0.000677)(0.340000,0.999900)+-(0.000438,0.000438)(0.345000,0.999380)+-(0.005011,0.005011)(0.350000,0.999660)+-(0.000807,0.000807)(0.355000,0.999500)+-(0.001150,0.001150)(0.360000,0.999120)+-(0.002280,0.002280)(0.365000,0.998880)+-(0.003859,0.003859)(0.370000,0.998420)+-(0.003319,0.003319)(0.375000,0.998060)+-(0.006706,0.006706)(0.380000,0.996560)+-(0.009416,0.009416)(0.385000,0.995120)+-(0.008137,0.008137)(0.390000,0.993760)+-(0.010773,0.010773)(0.395000,0.995740)+-(0.000787,0.000787)(0.400000,0.990220)+-(0.003855,0.003855)(0.405000,0.987480)+-(0.001123,0.001123)(0.410000,0.986300)+-(0.001078,0.001078)(0.415000,0.984980)+-(0.022157,0.022157)(0.420000,0.969680)+-(0.030926,0.030926)(0.425000,0.977180)+-(0.001844,0.001844)(0.430000,0.955500)+-(0.019740,0.019740)(0.435000,0.944)+-(0.021129,0.021129)(0.440000,0.943)+-(0.025413,0.025413)(0.445000,0.91)+-(0.031353,0.031353)(0.450000,0.898400)+-(0.028518,0.028518)(0.455000,0.88)+-(0.044574,0.044574)(0.460000,0.85)+-(0.054132,0.054132)(0.465000,0.83)+-(0.032135,0.032135)(0.470000,0.81)+-(0.028108,0.028108)(0.475000,0.78)+-(0.047582,0.047582)(0.480000,0.74)+-(0.051884,0.051884)(0.485000,0.70)+-(0.061478,0.061478)(0.490000,0.68)+-(0.071408,0.071408)(0.495000,0.62)+-(0.070329,0.070329)(0.500000,0.58)+-(0.073515,0.073515)
};\label{ME}
\addplot[black, mark=x, mark size=2pt, smooth,error bars/.cd,y dir=both,y explicit, error bar style={mark size=.5pt}] plot coordinates{
(0.200000,1.000000)+-(0.000000,0.000000)(0.205000,1.000000)+-(0.000000,0.000000)(0.210000,1.000000)+-(0.000000,0.000000)(0.215000,1.000000)+-(0.000000,0.000000)(0.220000,1.000000)+-(0.000000,0.000000)(0.225000,1.000000)+-(0.000000,0.000000)(0.230000,1.000000)+-(0.000000,0.000000)(0.235000,1.000000)+-(0.000000,0.000000)(0.240000,1.000000)+-(0.000000,0.000000)(0.245000,1.000000)+-(0.000000,0.000000)(0.250000,0.999980)+-(0.000200,0.000200)(0.255000,0.999980)+-(0.000200,0.000200)(0.260000,1.000000)+-(0.000000,0.000000)(0.265000,0.999920)+-(0.000394,0.000394)(0.270000,0.999960)+-(0.000281,0.000281)(0.275000,0.999880)+-(0.000477,0.000477)(0.280000,0.999900)+-(0.000522,0.000522)(0.285000,0.999700)+-(0.000772,0.000772)(0.290000,0.999600)+-(0.001101,0.001101)(0.295000,0.996120)+-(0.024546,0.024546)(0.300000,0.993960)+-(0.035847,0.035847)(0.305000,0.996720)+-(0.017619,0.017619)(0.310000,0.992980)+-(0.027727,0.027727)(0.315000,0.995060)+-(0.014395,0.014395)(0.320000,0.990520)+-(0.022639,0.022639)(0.325000,0.987660)+-(0.027937,0.027937)(0.330000,0.978320)+-(0.046275,0.046275)(0.335000,0.961400)+-(0.090093,0.090093)(0.340000,0.964680)+-(0.068804,0.068804)(0.345000,0.973700)+-(0.063126,0.063126)(0.350000,0.992580)+-(0.034041,0.034041)(0.355000,0.996840)+-(0.023262,0.023262)(0.360000,0.999940)+-(0.000343,0.000343)(0.365000,0.999940)+-(0.000343,0.000343)(0.370000,0.999880)+-(0.000477,0.000477)(0.375000,0.999900)+-(0.000438,0.000438)(0.380000,0.997940)+-(0.019796,0.019796)(0.385000,0.998440)+-(0.013596,0.013596)(0.390000,0.999680)+-(0.000790,0.000790)(0.395000,0.999740)+-(0.000787,0.000787)(0.400000,0.999220)+-(0.003855,0.003855)(0.405000,0.999480)+-(0.001123,0.001123)(0.410000,0.999300)+-(0.001078,0.001078)(0.415000,0.996980)+-(0.022157,0.022157)(0.420000,0.993680)+-(0.030926,0.030926)(0.425000,0.998180)+-(0.001844,0.001844)(0.430000,0.995500)+-(0.019740,0.019740)(0.435000,0.995020)+-(0.021129,0.021129)(0.440000,0.992840)+-(0.025413,0.025413)(0.445000,0.988940)+-(0.031353,0.031353)(0.450000,0.988400)+-(0.028518,0.028518)(0.455000,0.983180)+-(0.044574,0.044574)(0.460000,0.974120)+-(0.054132,0.054132)(0.465000,0.980980)+-(0.032135,0.032135)(0.470000,0.980560)+-(0.028108,0.028108)(0.475000,0.965420)+-(0.047582,0.047582)(0.480000,0.956660)+-(0.051884,0.051884)(0.485000,0.936300)+-(0.061478,0.061478)(0.490000,0.916640)+-(0.071408,0.071408)(0.495000,0.874020)+-(0.070329,0.070329)(0.500000,0.856560)+-(0.073515,0.073515)
}; \label{Joint mean field}
		   \coordinate (top) at (rel axis cs:0,1);
        \nextgroupplot[xmajorgrids,ymajorgrids,
               enlarge x limits=false, xmin=0.3,
      ymin=0.2,
      xmax=0.5,
      ymax=1,
      grid=major,
      scaled ticks=true,
      xlabel={$q'$}]
                \addplot[blue, mark=triangle, mark size=2pt, smooth,error bars/.cd,y dir=both,y explicit, error bar style={mark size=.5pt}] plot coordinates{
(0.200000,0.990620)+-(0.016713,0.016713)(0.205000,0.991220)+-(0.007515,0.007515)(0.210000,0.988160)+-(0.016944,0.016944)(0.215000,0.985500)+-(0.014017,0.014017)(0.220000,0.987620)+-(0.021609,0.021609)(0.225000,0.984120)+-(0.016918,0.016918)(0.230000,0.981320)+-(0.018539,0.018539)(0.235000,0.981960)+-(0.021634,0.021634)(0.240000,0.981200)+-(0.023010,0.023010)(0.245000,0.978480)+-(0.019713,0.019713)(0.250000,0.970160)+-(0.032824,0.032824)(0.255000,0.969460)+-(0.027645,0.027645)(0.260000,0.970980)+-(0.027853,0.027853)(0.265000,0.971140)+-(0.022836,0.022836)(0.270000,0.968060)+-(0.029861,0.029861)(0.275000,0.961880)+-(0.024695,0.024695)(0.280000,0.956100)+-(0.031118,0.031118)(0.285000,0.949420)+-(0.040796,0.040796)(0.290000,0.944420)+-(0.040639,0.040639)(0.295000,0.943300)+-(0.038446,0.038446)(0.300000,0.944160)+-(0.031539,0.031539)(0.305000,0.930540)+-(0.038832,0.038832)(0.310000,0.934680)+-(0.031731,0.031731)(0.315000,0.924240)+-(0.042462,0.042462)(0.320000,0.920620)+-(0.036521,0.036521)(0.325000,0.918180)+-(0.037757,0.037757)(0.330000,0.906680)+-(0.040514,0.040514)(0.335000,0.902560)+-(0.041637,0.041637)(0.340000,0.895920)+-(0.040029,0.040029)(0.345000,0.886080)+-(0.041537,0.041537)(0.350000,0.878860)+-(0.042843,0.042843)(0.355000,0.878200)+-(0.038797,0.038797)(0.360000,0.857680)+-(0.043397,0.043397)(0.365000,0.857820)+-(0.041403,0.041403)(0.370000,0.842560)+-(0.036330,0.036330)(0.375000,0.838920)+-(0.035603,0.035603)(0.380000,0.817740)+-(0.045272,0.045272)(0.385000,0.809420)+-(0.043873,0.043873)(0.390000,0.799160)+-(0.038263,0.038263)(0.395000,0.790300)+-(0.037160,0.037160)(0.400000,0.767060)+-(0.044871,0.044871)(0.405000,0.748980)+-(0.039526,0.039526)(0.410000,0.738960)+-(0.038314,0.038314)(0.415000,0.711380)+-(0.054324,0.054324)(0.420000,0.692460)+-(0.062998,0.062998)(0.425000,0.650540)+-(0.090390,0.090390)(0.430000,0.600500)+-(0.109554,0.109554)(0.435000,0.566200)+-(0.107060,0.107060)(0.440000,0.514360)+-(0.105361,0.105361)(0.445000,0.467460)+-(0.089980,0.089980)(0.450000,0.433420)+-(0.074345,0.074345)(0.455000,0.412620)+-(0.062745,0.062745)(0.460000,0.391180)+-(0.057525,0.057525)(0.465000,0.358020)+-(0.045584,0.045584)(0.470000,0.345980)+-(0.046555,0.046555)(0.475000,0.324740)+-(0.043853,0.043853)(0.480000,0.313820)+-(0.039347,0.039347)(0.485000,0.304380)+-(0.037275,0.037275)(0.490000,0.292540)+-(0.025726,0.025726)(0.495000,0.289880)+-(0.020153,0.020153)(0.500000,0.287660)+-(0.019830,0.019830)
};\label{Spectral clustering}
\addplot[red, mark=square, mark size=2pt, smooth,error bars/.cd,y dir=both,y explicit, error bar style={mark size=.5pt}] plot coordinates{
(0.200000,0.999880)+-(0.001200,0.001200)(0.205000,1.000000)+-(0.000000,0.000000)(0.210000,0.999520)+-(0.004800,0.004800)(0.215000,0.999980)+-(0.000200,0.000200)(0.220000,0.998780)+-(0.011799,0.011799)(0.225000,0.999960)+-(0.000400,0.000400)(0.230000,0.999860)+-(0.000910,0.000910)(0.235000,0.999360)+-(0.005408,0.005408)(0.240000,0.997720)+-(0.020808,0.020808)(0.245000,0.999280)+-(0.005246,0.005246)(0.250000,0.997060)+-(0.015296,0.015296)(0.255000,0.998500)+-(0.007029,0.007029)(0.260000,0.998040)+-(0.013033,0.013033)(0.265000,0.998960)+-(0.006748,0.006748)(0.270000,0.996080)+-(0.025518,0.025518)(0.275000,0.998440)+-(0.004527,0.004527)(0.280000,0.995480)+-(0.019407,0.019407)(0.285000,0.991560)+-(0.032385,0.032385)(0.290000,0.989880)+-(0.033043,0.033043)(0.295000,0.990800)+-(0.028303,0.028303)(0.300000,0.993380)+-(0.017393,0.017393)(0.305000,0.987400)+-(0.024334,0.024334)(0.310000,0.990300)+-(0.016724,0.016724)(0.315000,0.982080)+-(0.034776,0.034776)(0.320000,0.982960)+-(0.030044,0.030044)(0.325000,0.980920)+-(0.034850,0.034850)(0.330000,0.972000)+-(0.038796,0.038796)(0.335000,0.969780)+-(0.044435,0.044435)(0.340000,0.965140)+-(0.041034,0.041034)(0.345000,0.956900)+-(0.047107,0.047107)(0.350000,0.948540)+-(0.051307,0.051307)(0.355000,0.949380)+-(0.047497,0.047497)(0.360000,0.929020)+-(0.054980,0.054980)(0.365000,0.929480)+-(0.053234,0.053234)(0.370000,0.915720)+-(0.045629,0.045629)(0.375000,0.913480)+-(0.045616,0.045616)(0.380000,0.888860)+-(0.058835,0.058835)(0.385000,0.878420)+-(0.054997,0.054997)(0.390000,0.865560)+-(0.052097,0.052097)(0.395000,0.854960)+-(0.049182,0.049182)(0.400000,0.829060)+-(0.053522,0.053522)(0.405000,0.806240)+-(0.051781,0.051781)(0.410000,0.797160)+-(0.046655,0.046655)(0.415000,0.758900)+-(0.070653,0.070653)(0.420000,0.738580)+-(0.077688,0.077688)(0.425000,0.687940)+-(0.109746,0.109746)(0.430000,0.625200)+-(0.127601,0.127601)(0.435000,0.586060)+-(0.124483,0.124483)(0.440000,0.526560)+-(0.124598,0.124598)(0.445000,0.469020)+-(0.105926,0.105926)(0.450000,0.430300)+-(0.087231,0.087231)(0.455000,0.408280)+-(0.070995,0.070995)(0.460000,0.381400)+-(0.061396,0.061396)(0.465000,0.350960)+-(0.042681,0.042681)(0.470000,0.336400)+-(0.041108,0.041108)(0.475000,0.316440)+-(0.039705,0.039705)(0.480000,0.311460)+-(0.034581,0.034581)(0.485000,0.302800)+-(0.030231,0.030231)(0.490000,0.290340)+-(0.022720,0.022720)(0.495000,0.290280)+-(0.020392,0.020392)(0.500000,0.287940)+-(0.018537,0.018537)
};\label{Independent mean field}
\addplot[black, mark=x, mark size=2pt, smooth,error bars/.cd,y dir=both,y explicit, error bar style={mark size=.5pt}] plot coordinates{
(0.200000,1.000000)+-(0.000000,0.000000)(0.205000,1.000000)+-(0.000000,0.000000)(0.210000,1.000000)+-(0.000000,0.000000)(0.215000,1.000000)+-(0.000000,0.000000)(0.220000,0.997640)+-(0.023600,0.023600)(0.225000,0.999360)+-(0.006400,0.006400)(0.230000,1.000000)+-(0.000000,0.000000)(0.235000,0.999940)+-(0.000343,0.000343)(0.240000,0.997760)+-(0.022400,0.022400)(0.245000,0.999900)+-(0.000522,0.000522)(0.250000,0.993780)+-(0.034958,0.034958)(0.255000,0.997140)+-(0.022389,0.022389)(0.260000,0.995100)+-(0.028955,0.028955)(0.265000,0.990240)+-(0.040348,0.040348)(0.270000,0.985000)+-(0.048926,0.048926)(0.275000,0.970100)+-(0.077101,0.077101)(0.280000,0.966080)+-(0.083787,0.083787)(0.285000,0.962620)+-(0.088595,0.088595)(0.290000,0.958060)+-(0.082711,0.082711)(0.295000,0.966360)+-(0.061276,0.061276)(0.300000,0.966120)+-(0.071268,0.071268)(0.305000,0.957280)+-(0.085760,0.085760)(0.310000,0.955740)+-(0.085868,0.085868)(0.315000,0.966780)+-(0.067586,0.067586)(0.320000,0.978520)+-(0.053196,0.053196)(0.325000,0.980980)+-(0.053715,0.053715)(0.330000,0.984780)+-(0.050050,0.050050)(0.335000,0.996180)+-(0.014809,0.014809)(0.340000,0.993760)+-(0.022486,0.022486)(0.345000,0.993880)+-(0.028284,0.028284)(0.350000,0.993120)+-(0.028682,0.028682)(0.355000,0.996100)+-(0.021102,0.021102)(0.360000,0.989380)+-(0.035656,0.035656)(0.365000,0.990440)+-(0.031941,0.031941)(0.370000,0.987140)+-(0.040606,0.040606)(0.375000,0.984380)+-(0.043655,0.043655)(0.380000,0.979800)+-(0.048188,0.048188)(0.385000,0.986400)+-(0.037520,0.037520)(0.390000,0.984480)+-(0.032909,0.032909)(0.395000,0.979120)+-(0.043315,0.043315)(0.400000,0.974380)+-(0.049248,0.049248)(0.405000,0.961920)+-(0.064156,0.064156)(0.410000,0.956820)+-(0.063457,0.063457)(0.415000,0.954740)+-(0.058776,0.058776)(0.420000,0.950140)+-(0.060371,0.060371)(0.425000,0.935000)+-(0.067453,0.067453)(0.430000,0.912380)+-(0.073032,0.073032)(0.435000,0.910640)+-(0.067186,0.067186)(0.440000,0.884280)+-(0.070565,0.070565)(0.445000,0.879640)+-(0.067689,0.067689)(0.450000,0.867400)+-(0.064841,0.064841)(0.455000,0.851680)+-(0.057199,0.057199)(0.460000,0.841040)+-(0.055229,0.055229)(0.465000,0.822460)+-(0.045672,0.045672)(0.470000,0.819220)+-(0.048405,0.048405)(0.475000,0.809600)+-(0.038026,0.038026)(0.480000,0.799160)+-(0.032319,0.032319)(0.485000,0.797240)+-(0.030934,0.030934)(0.490000,0.790160)+-(0.022877,0.022877)(0.495000,0.786660)+-(0.020408,0.020408)(0.500000,0.789100)+-(0.016184,0.016184)
}; \label{Joint mean field} 
\coordinate (bot) at (rel axis cs:1,0);
    \end{groupplot}
 legend
\path (myplot c1r1.north west|-current bounding box.north)--
      coordinate(legendpos)
      (myplot c2r1.north east|-current bounding box.north);
\matrix[
  matrix of nodes,
  anchor=south,
   draw,
   inner sep=0.2em,
   draw
  ]at([yshift=1ex]legendpos)
  {
    \ref{Spectral clustering}& SC &[2pt]
    \ref{Independent mean field}& Single MF &[2pt]
     \ref{ME}& M-E &[2pt]
     \ref{Joint mean field}& Joint MF \\};
\end{tikzpicture}
 \caption{Normalized Mutual Information (NMI) between communities (of noisier graph $ \mathcal{G}^{{(2)}}$) identified by different community detection algorithms and ground truths, $ n=500, $ $ K=2 $ shared communities between the two graphs, $ K^{(l)}=4.$ Averages over $ 100 $ randomly generated graphs.}
  \label{fig:perf2}
\end{figure}
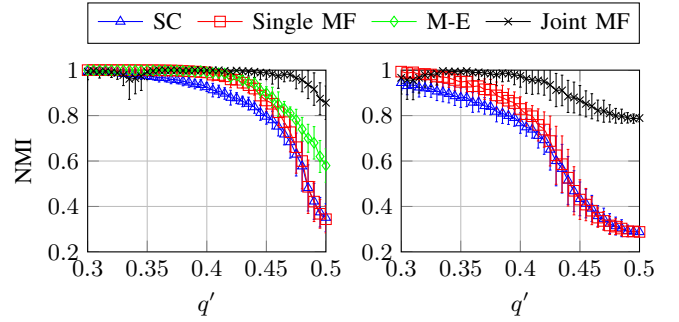

We first consider two ${\rm Bernoulli}$ SBM graphs $ \mathcal{G}^{{(1)}} $ and $ \mathcal{G}^{{(2)}} $ with the same intra-community probabilities and different inter-community probabilities in such a way that one graph is noisier than the other. Blindly identifying the community labels from each of the graphs would yield poor performances since we do not know in advance which graph has a clearer community structure than the other. $ \mathcal{G}^{{(1)}} $ and $ \mathcal{G}^{{(2)}} $ are constructed with $ n=500 $ vertices, each partitioned into $ K^{(l)}=4 $ ($ l=1,2 $) communities respectively among which $ K=2 $ are shared between the two graphs. The community labels $ g^{(l)}_i $ are assigned uniformly over the intervals $ \{1,\ldots,K^{(l)}\} $ in such a way that $ g^{(1)}_i=g^{(2)}_i $ when $ g^{(1)}_i \in \{1,\ldots,K\}.$ Given the community labels $ g^{(l)}_i,$ the entries of the adjacency matrices are generated as $ A^{(l)}_{ij} \sim {\rm Bernoulli} ({\bm \theta}^{(l)}) $ with $ {\bm \theta}^{(1)}=(p-q){\bm I}_4+q{\bf 1}_4{\bf 1}_4^{\sf T} $ and $ {\bm \theta}^{(2)}=(p-q'){\bm I}_4+q'{\bf 1}_4{\bf 1}_4^{\sf T},$ where we fix $ p=0.6, $ $ q=0.2 $ and we vary $ q' $ between $ 0.2 $ and $0.5.$ The larger $ q' $ is, the noisier the graph $ \mathcal{G}^{{(2)}}$ is in comparison with $ \mathcal{G}^{{(1)}}$ and the more difficult community recovery is when applying a community detection algorithm on $ \mathcal{G}^{{(2)}} $ solely. The left figure in Figure~\ref{fig:perf2} shows that our Joint mean field (MF) algorithm outperforms the competing method Multilayer Extraction~\cite{wilson2017community} (M-E) in extracting both shared and unshared communities from the two correlated graphs. Both methods are designed to exploit the graph $ \mathcal{G}^{{(1)}} $ to identify the communities of $ \mathcal{G}^{{(2)}} $ (among which some are shared with $ \mathcal{G}^{{(1)}} $). Both joint methods significantly outperform a spectral clustering (SC) algorithm and a mean field variational algorithm applied on $ \mathcal{G}^{{(2)}}$ alone. 

We next consider the same graph settings as before but with disparate distributions $ A^{(1)}_{ij} \sim {\rm Bernoulli}({\bm \theta}^{(1)}) $ and $A^{(2)}_{ij}\sim {\rm Poisson} ({\bm \theta}^{(2)}).$ Here the M-E algorithm is not exploitable since the latter is designed only for binary graphs (${\rm Bernoulli}$). Our joint variational algorithm is thus compared with single-graph clustering algorithms. The results are reported in the right figure of Figure~\ref{fig:perf2} where our joint algorithm outperforms single-layer clustering algorithms (spectral clustering and mean field variational Bayes).

\subsection{Real world graphs}
In this section, we make use of our novel approach to understand the interplay between genome structure (form) and   transcription (function) based on
a human fibroblast proliferation dataset \cite{chen2015functional}. 
This dataset consists of Hi-C contact maps \cite{LiebermanAiden09} that capture chromatin architectures and RNA-seq data that provide gene expression levels over $8$ time points.
For each chromosome ($ 1-22 $), we first build a correlation matrix between the RNA-seq values, where thresholding is applied to obtain a binary adjacency matrix $ {\bf A}^{(1)} $ representing functional correspondence between different genes. The threshold was chosen as the mean of the entries of the correlation matrix.
We then construct an average (over the $ 8 $ time points) Hi-C matrix $ {\bf A}^{(2)} $ and round each entry of the average matrix to the closest integer value. For the application of the variational Bayes algorithm, the entries of $ {\bf A}^{(1)} $ are considered to be ${\rm Bernoulli}$ distributed while those of $ {\bf A}^{(2)} $ are considered to be ${\rm Poisson}$ distributed. More sophisticated models for the sample correlation graph, e.g., Wishart distributions, could also be considered but this is left for future work.   

\begin{figure}
\centering
\includegraphics[width=.85\linewidth, height=.65\linewidth]{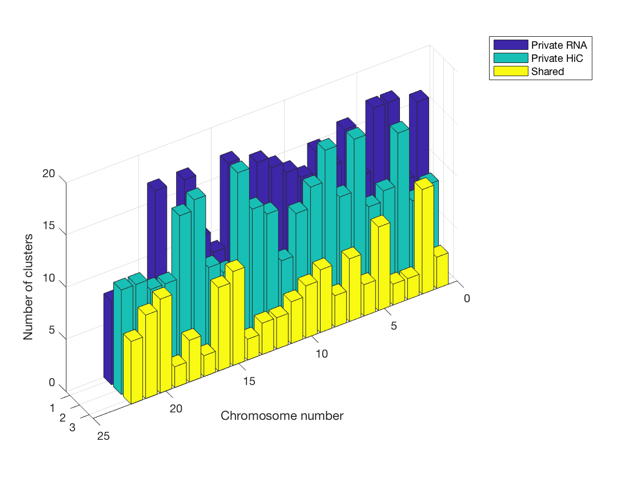}
 \caption{Histogram of the number of clusters (private and shared) for each chromosome.}
  \label{fig:histogram}
\end{figure}

We make use of our joint variational algorithm to automatically determine the number of shared clusters between the two graphs ($ {\bf A}^{(1)} $ and $ {\bf A}^{(2)} $) as well as the number of clusters unique to each graph. We first determine the total number of clusters $ K^{(1)},$ $ K^{(2)}$ respectively for each graph by applying separately the single layer variational Bayes community detection algorithm~\cite{aicher2014learning} and choose the number of clusters maximizing the Bayesian Information Criteria (BIC) ~\cite{yan2016bayesian}. We then apply our joint method with fixed $ K^{(1)} $ and $ K^{(2)}$ (determined previously) and choosing the number of shared clusters $ K $ (in the range $ [2,\min(K^{(1)},K^{(2)})] $) maximizing the modularity~\cite{newman2006modularity} in both graphs. Figure~\ref{fig:histogram} shows for each chromosome the number of shared clusters, the number of private clusters for graph $ {\bf A}^{(1)} $ (Private RNA) and the number of private clusters for graph $ {\bf A}^{(2)} $ (Private HiC). This analysis suggests that the genes in chromosomes $ 2, 5, 15, 16, 20, 21, 22 $ are strongly co-expressed (high connectivity in $ {\bf A}^{(1)} $) and strongly connected (high connectivity in $ {\bf A}^{(2)} $) since the number of shared clusters between graphs $ {\bf A}^{(1)} $ and $ {\bf A}^{(2)} $ is dominant compared to the number of private clusters, while the genes in the other chromosomes are either strongly co-expressed or strongly connected.

\section{Conclusion}
Our proposed joint variational algorithm is capable of extracting shared communities across all graph layers as well as identifying communities unique to each layer. The method is applicable to any multilayer network (with or without edge weights) and can provide important insights in the analysis of real-world systems as demonstrated for the human fibroblast dataset. An interesting direction of future investigation would be to consider extensions to the case where only a subset of layers share communities.
\subsubsection*{{\bf Acknowledgments}}
The authors thank Zeyu Sun at the University of Michigan for his help implementing the numerical experiments.

\bibliographystyle{IEEEbib}
\bibliography{biblio1}

\end{document}